\def\arcm     {{\ifmmode {'\ }\else$'     $\fi} } 
\def\arcs     {{\ifmmode{''\ }\else$''    $\fi} } 

\def\Msun     {{\ $M_{\odot}$} }

\def\Msun     {{\ $M_{\odot}$} }

\documentstyle[epsfig,longtable]{aipproc}

\begin{document}

\title{A Radio Perspective on Star-Formation in Distant Galaxies}
 
\author{Eric A. Richards}
\address{National Radio Astronomy Observatory \& University of Virginia \\  520
Edgemont Road \\ Charlottesville, VA 22903\\
email: erichard@nrao.edu}

\lefthead{Radio Star-Formation}
\righthead{Richards}
\maketitle

\vspace*{-8mm}
\begin{abstract}

        Determination of the epoch dependent star-formation
rate of field galaxies is one of the principal goals
of modern observational cosmology. Deep radio
surveys, sensitive to starbursts out to
$z \sim$ 1-2, may hold the key to understanding
the evolution of the starburst phenonemon
unhindered by the effects of dust. Using
deep, high resolution radio observations of the
Hubble Deep Field, we show that the $\mu$Jy radio 
emission from field galaxies at $z\sim 0.4-1$ 
is primarily starburst in origin. In addition,
we have discovered a population of optically
faint, possibly obscured systems that are candidate
high-$z$ protogalaxies. At least one of these radio
sources is identified with a sub-mm detection.

\end{abstract}

\vspace*{-6mm}
\section*{Radio Emission as a Star Formation Tracer}
\vspace*{-4mm}

         The diffuse radio emission observed in local
starbursts is believed to be a mixture of
synchrotron radiation (excited by supernovae
remnants and hence directly proportional
to the number of supernovae producing stars)
and thermal radiation (from HII regions and hence
an indicator of the number of O and B stars in a galaxy).
As the thermal and synchrotron radiation of a
starburst dissipates on a
physical time scale of $10^7-10^8$ years, the
radio luminosity is a true measure of the
instantaneous star-formation rate (SFR) in a galaxy, uncontaminated by
older stellar populations.  Since supernovae
progenitors are dominated by $\sim$8 \Msun stars,
synchrotron radiation has the additional advantage of being
less sensitive to uncertainties in the initial
mass function as opposed to UV and optical
recombination line emission. However, the most
obvious advantage of using the radio luminosity
as a SFR tracer is its unsusceptibility to dust
obscuration, as galaxies and the inter-galactic
medium are transparant at centimeter wavelengths.

	Comparison of the local radio
luminosity function (LF) of star-forming galaxies
\cite{C89}
with those derived independently at
FIR \cite{S87}, H$\alpha$\cite{G95}, and
UV  wavelengths\cite{T98} shows surprsing agreement.
Figure 1 shows the four
LFs in units of SFRs.
This plot suggests that the bulk
of local star formation is occurring in
modest starbursts with SFR $\sim$ 10 \Msun yr$^{-1}$.
However, past the peak in the
LF, the H$\alpha$ and UV estimates
begin to drop below the radio/FIR rates,
and beyond 50\Msun yr$^{-1}$ has
entirely vanished. {\bf This is direct evidence
that optically selected surveys are incapable
of detecting the most extreme
and dust obscured starbursts.}

\vspace*{-6mm}
\section*{Radio Emission from Distant Galaxies}
\vspace*{-4mm}

        Our 1.4/8.5 GHz study of the Hubble Deep Field using
the VLA and MERLIN
has demonstrated that 70\% of $\mu$Jy sources are
identified with morphologically peculiar, merging and/or interacting disk
galaxies, many with independent evidence of star-formation
(blue colors, infra-red excess, HII optical spectra) \cite{R98a}
\cite{R98b}.
Radio morphologies from the high
resolution VLA/MERLIN observations of the HDF \cite{M98}
indicate that
95\% of $\mu$Jy radio sources are resolved
at 0.2\arcsec ~resolution and suggests a
median size of 2\arcsec, comparable to
the optical extent of these $z \sim 0.4-1$ systems.
These data exclude AGN as the dominant contributor
to the radio luminosity in the majority of these
systems.
Thus the cosmological faint radio population
is dominated by the distant analogs
of local IRAS galaxies with suggested
star-formation rates of 10-1000 \Msun yr$^{-1}$.
 
\nopagebreak[4]
\vspace*{-0mm}
\begin{figure} 
\centerline{\epsfig{file=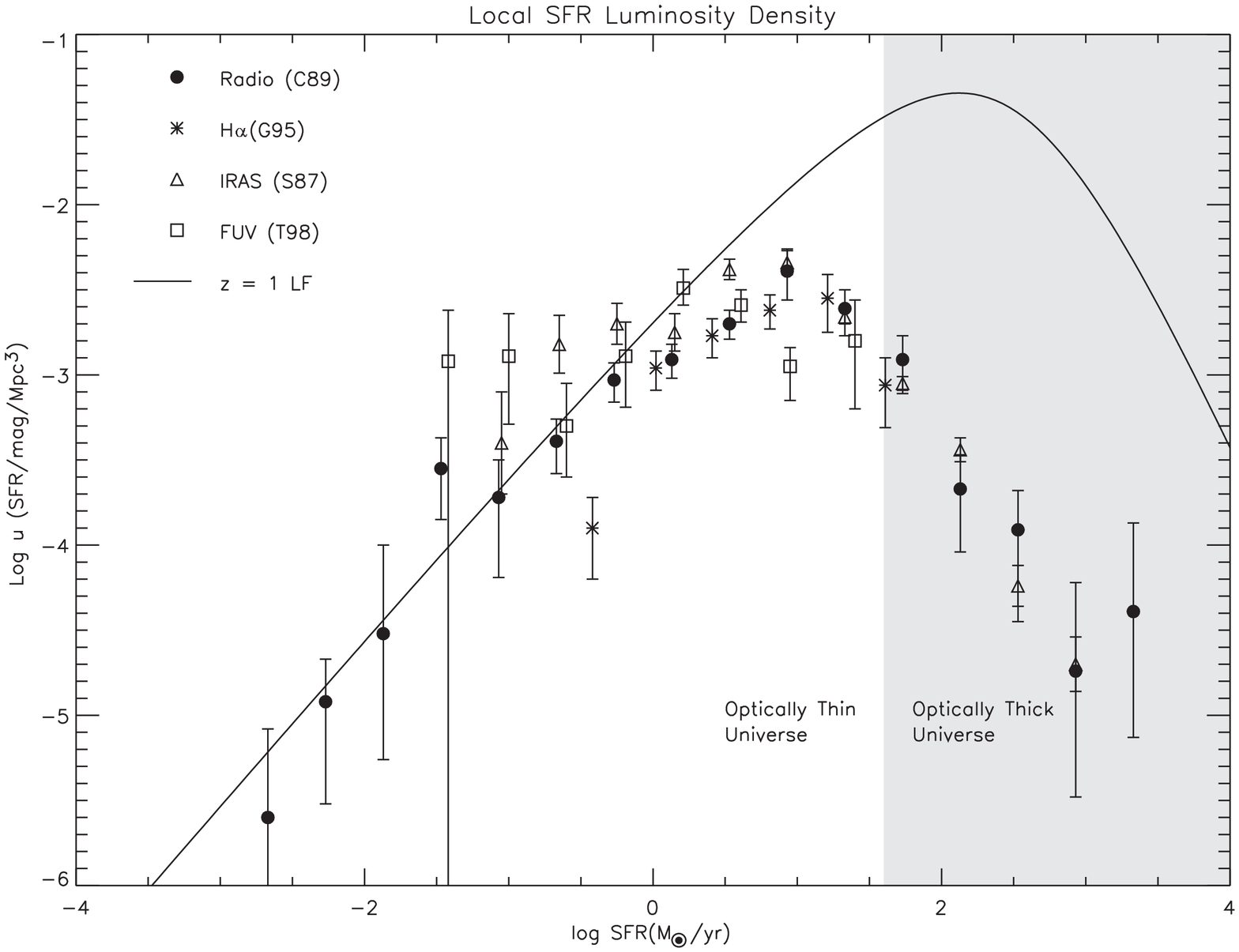,width=4.0in,height=3.5in}}
\vspace*{-4mm}
\caption{
Shown is the contribution to the local
star-formation luminosity density (u) per luminosity interval of
star-forming galaxy. The radio, IRAS, H$\alpha$, and
far-ultraviolet (FUV) luminosities have been converted to SFRs
assuming a Salpeter IMF over 0.1-100 \Msun .
Notice that the four
measures of SFRs all peak at
$\sim$ 10 \Msun yr$^{-1}$ (see Cramm, ApJL, 506, 85 for a fuller
discussion).
The radio and IRAS points are in particularly good agreement,
reflecting the tight FIR/radio correlation in
star-forming galaxies.
The shaded region represents
what may be a dust curtain beyond which optical
surveys are blind to star formation.
If the SFR luminosity function
evolves as L $\propto (1+z)^{3.5}$, then by $z$ = 1,
it will appear as the
solid line. This analysis suggests that the
bulk of global star-formation at high$-z$ is hidden from optical
surveys.}
\end{figure}
\nopagebreak[4]

\vspace{-0mm}

\begin{figure}[htb] %
\centerline{\epsfig{file=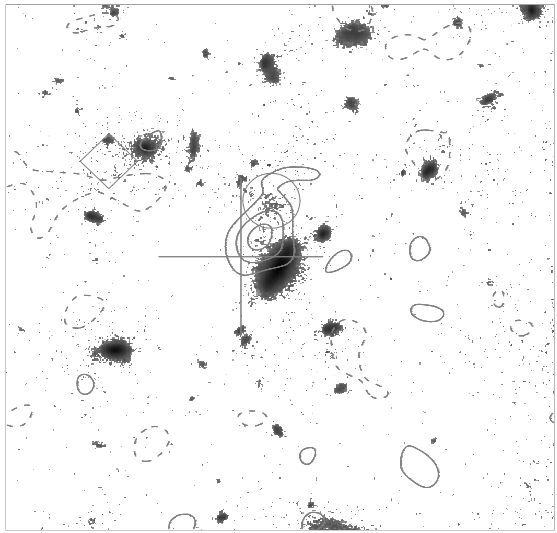,width=3.0in,height=3.0in}}
\vspace*{-0mm}
\caption{
The greyscale shows a 20\arcsec $\times$ 20\arcsec ~HDF
I-band image containing the
SCUBA detection HDF850.1. The contours correspond to
1.4 GHz emission at the -2, 2, 4 and 6 $\sigma$ level ($\sigma$ = 7.5 $\mu$Jy).
The three sigma position error circle for HDF850.1 is shown
after shifting to the VLA coordinate frame. The original
position of HDF850.1 taken from Hughes et al. (1998) is
denoted by the diamond. The ISO detection is marked with
a cross with three sigma position errors (Aussel et al. 1998,
A \& A in press).
The 0.1\arcsec ~radio/optical registration
clearly rules out association with the bright spiral.
VLA3649+1221 may be the most obscured part of
a larger galaxy 3-633.1 at $z$ = 1.72 
(located
directly underneath the SCUBA error circle; Fernandez-Soto
et al. 1998, AJ in press).
}

\end{figure}

	We have detected over 100 radio sources
in complete samples within 4.5\arcmin ~of the
HDF where deep optical imaging is available.
Ninety percent of these radio sources are
identified with galaxies of mean magnitude
I$_{AB}\sim$ 22. However, approximately 10\%
remain unidentified to I$_{AB}$ = 28 in the
HDF and I$_{AB}$ = 26 in the HDF flanking
fields. NICMOS imaging of one of these unidentifed
radio sources (VLA3642+1331) has revealed a H = 22.7 
{\em disk} galaxy at unknown redshift (see I.
Waddington elsewhere in these proceedings).
Another radio source, VLA3649+1221,
is identified with a $I_{AB}$ = 28
object in the HDF. NICMOS imaging
by Dickinson et al. (1998, private communication)
shows that this object has a steeply rising 
optical/near infrared spectrum possibly
suggesting a high redshift.
We have subsequently shown that the
brightest sub-mm/SCUBA source in the HDF
(HDF850.1) (Hughes et al. 1998, Nature, 393, 241) is identified
with VLA3649+1221. 
{\em It is our hypothesis that 
these systems are extreme,
dust obscured starburst galaxies.} 
The surface density of
these candidate high-$z$ radio objects is
comparable to the sub-mm population at $\sim$
0.1 arcmin$^{-2}$.

\vspace{0mm}

\begin{figure}[htb] %
\centerline{\epsfig{file=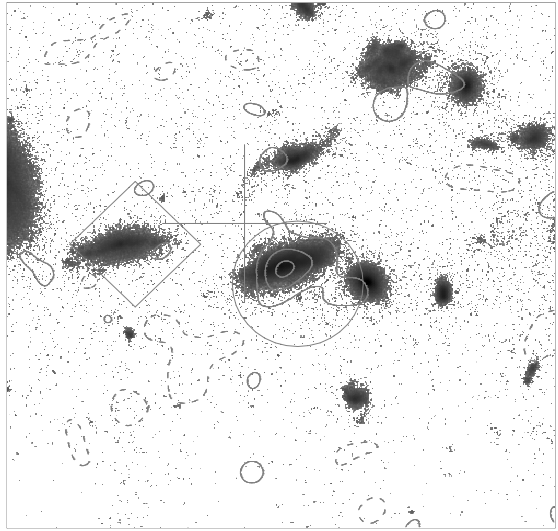,width=3.0in,height=3.0in}}
\vspace{0pt}
\caption{
Radio 1.4 GHz contours drawn at the
-2, 2, 4 and 6 $\sigma$ level are overlaid on the
HDF I-band image centerd on the position of
 HDF850.4 (20\arcsec ~on a side). A 15 $\mu$m detection
 from the complete catalog of
Aussel et al. (1998) has been associated with this
radio source and suggests likely starburst
activity in the disk galaxy. The symbols are the
same as for Figure 2. We estimate a SFR = 150\Msun yr$^{-1}$.}
\end{figure}

        Of the five 850 $\mu$m sources in the HDF,
two are solidly detected at radio wavelengths (Fig.
2 and 3), while two have less secure radio 
identifications\cite{R98c}. 
Two of these radio/sub-mm sources are
associated with
$z \sim$ 0.5
starbursts (HDF850.3 and HDF850.4). The other
two detections (HDF850.1 and HDF850.2) must have
redshifts less than 1.5 or be contaminated by
AGN based on radio luminosity arguments.
This radio analysis suggests  the
claim, based on low resolution sub-mm observations alone,
that the optical surveys underestimate the
$z>2$ global star-formation rate are premature.
On the other hand, the $z < 1$ star-formation
history may have been underestimated if a significant
fraction of the sub-mm population lies at relatively
low redshift, in agreement with the large number
of $z$ = 0.4-1 radio starbursts we are finding 
in the HDF. Analysis of the evolving radio star-forming
luminosity function promises to shed light on 
the prevalence of dust enshrouded starbrust activity
at high$-z$ (see Haarsma \& Partridge, elsewhere in
these proceedings).

\vspace{1mm}
\setlength{\baselineskip}{8pt}

{
\small
	It is a pleasure to thank my collaborators
K. Kellermann, E. Fomalont, B. Partridge, 
R. Windhorst, T. Muxlow,  I. Waddington, and D. Haarsma.
Support for part of
this work was provided by NASA through grant AR-6337.*-96A from the
STSI, which is operated by AURA,
Inc., under NASA contract
NAS5-2655, and by NSF grant AST 93-20049.
}


\vspace{-4mm}

\begin{references}

\vspace{-2mm}


\bibitem{C89}
Condon, J. J. 1989, ApJ 338, 13

\bibitem{S87}
Soifer, B. T. et al. 1987, ApJ, 320, 238

\bibitem{G95}
Gallego, J., Zamorano, J., Aragon-Salamanca, A. \& Rego, M. 1995,ApJL, 459, 43

\bibitem{T98}
Treyer, M., Ellis, R., Milliard, B., Donas, J. \& Bridges, T. 1998, MNRAS, in press

\bibitem{R98a}
Richards, E. A., et al. 1998, AJ, 116, 1039

\bibitem{R98b}
Richards, E. A. 1998, in preparation

\bibitem{M98}
Muxlow, T. W. et al., in preparation




\bibitem{R98c}
Richards, E. A. 1998, ApJL, submitted


\end{references}
\end{document}